\documentclass[12pt]{article}

\usepackage{amsmath}
\usepackage{graphicx}
\usepackage{xcolor}

\begin{document}

\title{A microscopic model for the self-inductance of an ideal solenoid}
\author{C\u{a}lin Galeriu}
\date{}

\maketitle

\section*{Abstract}

{\it We derive the formula for the self-inductance of an ideal solenoid by
calculating the total kinetic energy 
associated with the drift velocity of the conduction electrons.}

\section{Introduction}

From a mathematical point of view, 
the simple harmonic oscillator (with mass $m$ and elastic constant $k$)
and the LC oscillating circuit (with self-inductance $L$ and capacitance $C$)
are described by the same differential equations.
What happens in the mechanical system, 
the periodic transformation of the kinetic energy $m \, v^2 / 2 $ 
into the elastic potential energy $k \, \Delta x^2 / 2$,
and {\it vice versa}, also happens in the electrical system,
where the potential energy is the electric energy $Q^2 / (2 \, C)$ stored in the capacitor, 
and the kinetic energy is the magnetic energy $L \, I^2 / 2$ stored in the inductor.
Since the magnetic energy depends on the intensity $I$ of the electric current flowing
through the coil, which is due to the drift motion of the free electrons,
it is natural to assume that the magnetic energy stored in the inductor,
being a kinetic energy, must be the total kinetic energy 
associated with the drift velocity of the conduction electrons found in the coil.
According to E.~G.~Cullwick,
\lq\lq [...] the hypothesis is introduced that the magnetic energy of a current circuit 
is the same as the kinetic energy of the effective conductive electrons,
by virtue of their mean velocity along the wire.\rq\rq\ \cite{cullwick}
Based on this insight, we will derive the formula for the 
self-inductance $L$ of an ideal solenoid from a microscopic model.

\section{A review of electromagnetic induction}

The very existence of the electric potential energy seems to imply 
that the electric force is a conservative force, but this is true only for
a static distribution of source charges.
\lq\lq In general, the total $\vec{E}$ field at a point in space can be the
superposition of an electrostatic field $\vec{E}_c$ caused by a distribution of charges at
rest and a magnetically induced, nonelectrostatic field $\vec{E}_n$.\rq\rq\ \cite{zemansky}
This can be most clearly seen when we write the total electric field $\vec{E}$
in terms of the electromagnetic potentials
\begin{equation}
\vec{E} = - \nabla \phi - \frac{\partial \vec{A}}{\partial t}.
\label{eq1}
\end{equation}
The circulation of the conservative part of the electric field is zero
\begin{equation}
\oint \vec{E}_c \cdot \vec{dl} = - \oint \nabla \phi \cdot \vec{dl} = - \oint d \phi = 0,
\label{eq2}
\end{equation}
but the circulation of the nonconservative part of the electric field is equal to the induced electromotive force
\begin{equation}
\oint \vec{E}_n \cdot \vec{dl} = - \oint \frac{\partial \vec{A}}{\partial t} \cdot \vec{dl} 
= - \frac{d}{dt} \oint \vec{A} \cdot \vec{dl}
= - \frac{d}{dt} \int \vec{B} \cdot \vec{dS} = - \frac{d \Phi_{mgn}}{dt} = \mathcal{E},
\label{eq3}
\end{equation}
where first we have assumed that the coil is at rest, meaning that the integration path is stationary,
and then we have used Stoke's theorem.

Kirchhoff's loop rule, initially derived as a result of Eq. (\ref{eq2}), has to be carefully reconsidered when nonconservative
electric fields are present. The electric potential difference across an inductor is due to the conservative electric field that
is produced by \lq\lq accumulations of charge on the terminals of the inductor and the surfaces of its conductors \rq\rq \cite{zemansky},
electric charge distributions of whom very little is known. 
Another puzzling aspect is that these electric charge distributions
are obviously not constant in time, unless the electric current is also constant,
in which case the voltage across the inductor is zero.

A topic seldom discussed in physics textbooks is the fact that, for an ideal inductor with  zero electric resistance, 
\lq\lq the {\it total} electric field $\vec{E}_c + \vec{E}_n$ within the coils must be zero\rq\rq \cite{zemansky}.
\lq\lq The net electric field seen by the conduction electrons in the wire must be zero (assuming negligible
resistivity).\rq\rq\ \cite{heald}
For this reason $\vec{E}_c = - \vec{E}_n$ inside the wire and the voltage across the ideal inductor is calculated as
\begin{equation}
V_{ab} = V_a - V_b = \int_a^b \vec{E}_c \cdot \vec{dl} = - \int_a^b \vec{E}_n \cdot \vec{dl} = \frac{d \Phi_{mgn}}{dt} = L \frac{di}{dt},
\label{eq4}
\end{equation}
where the self-inductance $L$ shows the direct variation of the magnetic flux linkage $\Phi_{mgn}$ 
with the electric current of intensity $i$ through the inductor,
$\Phi_{mgn} = L \, i$.

The change $dU$ in the magnetic energy stored inside the ideal coil is equal to the electrical energy transferred from the voltage source.
Using Eq. (\ref{eq4}) we find that
\begin{equation}
dU = V_{ab} \, i \, dt = L \, i \, di,
\label{eq5}
\end{equation}
and the total magnetic energy $U$ is calculated as
\begin{equation}
U = \int_0^I L \, i \, di = \frac{L \, I^2}{2}.
\label{eq6}
\end{equation}

Suppose that we have an electric circuit with a voltaic battery, a resistor, and an
inductor connected in series. If we suddenly short-circuit the battery, the electric current
through the resistor will not stop instantly. 
The induced voltage in the coil will strongly oppose any 
abrupt change in the intensity of the 
electric current. The magnetic energy stored in the coil can still be used,
at least for a little while, to compensate for the ohmic losses in the resistor. 
But since this magnetic energy, according to E.~G.~Cullwick, is in fact the total kinetic 
energy associated with the drift velocity
of the conduction electrons in the coil, 
an alternative interpretation could be that
\lq\lq a current can be considered to flow in a short-circuited 
inductive circuit not because
of the induction of an e.m.f. but because of electron inertia.\rq\rq \cite{cullwick}

The self-inductance of a coil, like the mass of an object, 
is a measure of inertia, of opposition to change.
\lq\lq Whenever you try
to alter the current in a wire, you must fight against this back emf. Thus inductance plays
somewhat the same role in electric circuits that mass plays in mechanical systems: The
greater $L$ is, the harder it is to change the current, just as the larger the mass, the harder it
is to change an object's velocity.\rq\rq \cite{griffiths}
For this reason one would expect to see a relationship 
between the self-inductance of a coil 
and the mass of the conduction electrons, which are 
the free carriers of electric charge that move through the coil.
\lq\lq However, the self-inductance $L$ is known to be
independent of the electron’s mass $m$ and to depend
only on the geometry of the circuit.\rq\rq \cite{assis}
We will develop a microscopic model for the self-inductance of an ideal solenoid
that explains why the inertial mass of the electron
does not show up in the self-inductance formula.

\section{A microscopic model for self-inductance}

Consider an infinitely long air-core solenoid that is wrapped around the $z$ axis, with circular turns of radius $R$.
We focus our attention on a finite segment of this solenoid, a coil of length $\ell$ that has $N$ turns.
The infinite length of the full solenoid is needed in order to establish the translational symmetry 
of the system along the $z$ axis.
For the same reason we also assume that the wire is very thin, and that the loops of wire are wound very tightly in one layer,
such that we effectively work in the limit $\ell / N \to 0$. 
By definition an ideal solenoid is \lq\lq infinitely long and infinitely tightly wound\rq\rq \cite{mungan}.

The easiest way to determine the magnetic induction inside and outside an ideal coil involves the use of
symmetry, the Biot-Savart law, and Amp\`{e}re's law. 
\lq\lq The reflection of the solenoid in a mirror that is perpendicular to the solenoid axis 
does not change the direction of the current.\rq\rq \cite{hauser} 
The magnetic induction $\vec{B}$, being a pseudovector (an axial vector), 
is left unchanged by this symmetry operation only 
when it is parallel to the solenoid axis.
Once the axial direction of the magnetic induction 
has been established everywhere in space, $\vec{B} = (0, 0, B)$,
the Biot-Savart law is used to calculate the magnetic field $B_0$ on the solenoid axis \cite{mungan}. 
In this way we obtain the value
\begin{equation}
B_0 = \frac{\mu_0 \, N \, I}{\ell}.
\label{eq7}
\end{equation}
The use of the Biot-Savart law cannot be avoided in this particular situation, since we cannot simply
assume that the magnetic field vanishes very far away from the coil \cite{bunn}.
Finally, Amp\`{e}re's law is used with two rectangular integration contours, 
both of them with one side along the solenoid axis.
One integration path is completely inside the coil,
 but the other path is crossing the windings and reaching into the outside space.
The result is that everywhere inside the solenoid the magnetic induction has the constant value given by Eq. (\ref{eq7}),
$B_{in} = B_0$,
while everywhere on the outside of the solenoid the magnetic induction is zero, $B_{out} = 0$. 

Due to the zero electric resistance of the coil, as previously mentioned, the total electric field inside the wire is zero,
which means that only the magnetic part $\vec{F} = q \, \vec{v} \times \vec{B}$ 
of the Lorentz force acts upon the moving conduction electrons.
As a consequence, the trajectory of the electrons can be either circular or helicoidal \cite{orion}.
Since for a circular trajectory the electric charge is stuck in place and does not move along the $z$ axis,
as it should when an electric current flows through the solenoid, it is clear that the position coordinates
of a conduction electron are given by the equations
\begin{eqnarray}
x = R \cos(\omega \, t + \varphi_0), \label{eq8} \\
y = R  \sin(\omega \, t + \varphi_0), \label{eq9} \\
z = v_z \, t. \label{eq10}
\end{eqnarray}
The angular velocity $\omega$ is related to the period $T$ of the circular motion by the formula
\begin{equation}
\omega = \frac{2 \pi}{T}.
\label{eq11}
\end{equation}

From Eqs. (\ref{eq8})-(\ref{eq10}) the velocity components are obtained as
\begin{eqnarray}
v_x = - R \, \omega \sin(\omega \, t + \varphi_0), \label{eq12} \\
v_y = R \, \omega \cos(\omega \, t + \varphi_0), \label{eq13} \\
v_z = \textrm{constant}, \label{eq14}
\end{eqnarray}
and we notice that 
\begin{equation}
v_x^2 + v_y^2 = R^2 \, \omega^2.
\label{eq15}
\end{equation}

From Eqs. (\ref{eq12})-(\ref{eq14}) the acceleration components are obtained as
\begin{eqnarray}
a_x = - R \, \omega^2 \cos(\omega \, t + \varphi_0), \label{eq16} \\
a_y = - R \, \omega^2 \sin(\omega \, t + \varphi_0), \label{eq17} \\
a_z = 0. \label{eq18}
\end{eqnarray}

The components of the magnetic force $\vec{F} = q \, (v_x, v_y, v_z) \times (0, 0, B)$ are
\begin{eqnarray}
F_x = q \, v_y \, B = q \, B \, R \, \omega \cos(\omega \, t + \varphi_0), \label{eq19} \\
F_y = - q \, v_x \, B = q \, B \, R \, \omega \sin(\omega \, t + \varphi_0), \label{eq20} \\
F_z = 0, \label{eq21}
\end{eqnarray}
and Newton's second law, $\vec{F} = m \, \vec{a}$, reduces to
\begin{equation}
- q \, B = m \, \omega,
\label{eq22}
\end{equation}
where $m$ is the mass of an electron.
The electric charge $q$ of an electron is negative, and
we can write Eq. (\ref{eq22}) as
\begin{equation}
|q| \, B = m \, \omega.
\label{eq23}
\end{equation}

We are now faced with an interesting question: What is the value of the magnetic field $B$ at the position of the 
conduction electrons? We know that inside the solenoid the magnetic field is $B_0$, as given by Eq. (\ref{eq7}),
and that outside the solenoid the magnetic field is 0.
But what is the magnetic field $B$ on the boundary of these two spatial domains?
A first guess would be that we have to calculate the average of the $B_{in}$ and $B_{out}$ values, that is
\begin{equation}
B = \frac{\mu_0 \, N \, I}{2 \, \ell}.
\label{eq24}
\end{equation}
We can justify this step based on the fact that the electric and magnetic fields 
appearing in Maxwell's equations
are the {\it macroscopic} fields, produced by the averaging of the {\it microscopic} fields over some
very small volume elements \cite{jackson}.

Another, more mathematically sophisticated justification of Eq. (\ref{eq24}), relies on Amp\`{e}re's law, 
which has already been used
for obtaining the $B_{in}$ and $B_{out}$ values. 
We integrate the normal component of the electric current density over a surface bound by
the closed path along which the circulation of the magnetic field is calculated.
Consider a rectangular integration surface in the $xOz$ plane, with one side of the rectangle along the $Oz$ axis.
Each time a current carrying wire intersects this integration surface, lets say at $(R, 0, z_k)$, 
the electric current density shows up as a
2D Dirac delta function $\delta_k^{2D}(x, z) = \delta(x - R) \, \delta(z - z_k)$ 
in product with the electric current intensity $I$.
On one side of the solenoid, there are $N$ such intersections over a
distance $\ell$.
The integration of $\delta(z - z_k)$ along the length of the solenoid is uneventful, and gives 1,
but the integration of $\delta(x - R)$ splits into three cases. 
When the upper integration limit is less than $R$ we get 0, and this decides the value of $B_{in}$.
When the upper integration limit is greater than $R$ we get 1, and this decides the value of $B_{out}$.
But when the upper integration limit is equal to $R$ we get 1/2, 
as needed for the expression of $B$ from Eq. (\ref{eq24}).
This important property of the Dirac delta function was discussed by Amaku {\it et al.} \cite{Amaku2021RBEF}.

We obtain another equation by looking at the distance $\Delta \ell$ between two adjacent loops of wire in the 
solenoid. From geometrical considerations this distance is $\Delta \ell = \ell / N$. The same distance
must be travelled along the $z$ axis by one conduction electron in one period, which means that
\begin{equation}
v_z \, T = \frac{\ell}{N}.
\label{eq25}
\end{equation}

Let 
\begin{equation}
v = \sqrt{v_x^2 + v_y^2 + v_z^2}
\label{eq26}
\end{equation}
be the drift speed of each conduction electron that moves through the solenoid,
and let $n$ be the number of conduction electrons per unit length of wire, a linear density of particles.
The electric current intensity through the coil is
\begin{equation}
I = n \, v \, |q|.
\label{eq27}
\end{equation}

The length of one loop of wire in the solenoid is $v \, T$, 
the total length of the $N$ loops of wire is $N \, v \, T$,
and the total number of conduction electrons in the coil is $n \, N \, v \, T$.
The total kinetic energy 
associated with the drift velocity of the conduction electrons in the coil is
\begin{equation}
U = n \, N \, v \, T \frac{m \, v^2}{2}.
\label{eq28}
\end{equation}

Our goal is to write the kinetic energy $U$ from Eq. (\ref{eq28}) 
as an expression directly proportional to
the square of the electric current intensity, $U \sim I^2$.
In this way, 
by direct comparison with Eq. (\ref{eq6}),
we will extract the formula for the self-inductance $L$ of an ideal solenoid
in terms of the geometrical parameters of the coil.

Eq. (\ref{eq15}) gives us an expression for $v_x^2 + v_y^2$, and
from Eq. (\ref{eq25}) we obtain an expression for $v_z$.
These expressions are substituted into Eq. (\ref{eq26}), giving
\begin{equation}
v = \sqrt{R^2 \, \omega^2 + \dfrac{\ell^2}{N^2 \, T^2}}.
\label{eq29}
\end{equation}

From Eq. (\ref{eq27}) we obtain an expression for $n \, v$, which we substitute into Eq. (\ref{eq28}),
and then we substitute the remaining $v^2$ with its expression from Eq. (\ref{eq29}).
The kinetic energy becomes
\begin{equation}
U = \frac{I}{|q|} N \, T \frac{m}{2} \left(R^2 \, \omega^2 + \dfrac{\ell^2}{N^2 \, T^2}\right).
\label{eq30}
\end{equation}

From Eq. (\ref{eq11}) we obtain an expression for $T$, which we substitute into Eq. (\ref{eq30}).
The kinetic energy becomes
\begin{equation}
U = \frac{I}{|q|} N \, m \, \omega \left(\pi \, R^2 + \dfrac{\ell^2}{4 \, \pi \, N^2}\right).
\label{eq31}
\end{equation}

In the limit $\ell / N \to 0$ of an ideal solenoid, all that remains inside the round brackets is the 
surface area $S = \pi \, R^2$ of one turn of the coil.

We have now reached the moment when the mass $m$ of the electron disappears from the formula.
This is due to Eq. (\ref{eq23}) that describes the circular motion of an electrically charged particle in a 
magnetic field. The kinetic energy becomes
\begin{equation}
U = I \, N \, B \, S.
\label{eq32}
\end{equation}

As a side note,
this formula could also be written in terms of the magnetic flux linkage,
which is $\Phi_{mgn} = N \, B_{in} \, S = 2 \, N \, B \, S$, and thus $U = I \, \Phi_{mgn} / 2$.

In the final step we substitute the expression of the magnetic field $B$ from Eq. (\ref{eq24}) into Eq. (\ref{eq31}).
The kinetic energy becomes
\begin{equation}
U = \frac{\mu_0 \, N^2 \, S \, I^2}{2 \, \ell}.
\label{eq33}
\end{equation}

Direct comparison of Eqs. (\ref{eq6}) and (\ref{eq33}) gives us the correct formula for the
self-inductance of an ideal solenoid
\begin{equation}
L = \frac{\mu_0 \, N^2 \, S}{\ell}.
\label{eq34}
\end{equation}

The calculations presented here provide the definitive answer to the question
\lq\lq How does an inductor store magnetic energy?\rq\rq\ 
asked on the Physics Stack Exchange more than 10 years ago 
and summarily addressed by Bhupi (on Aug 5, 2023) without the full mathematical details:
\lq\lq Energy stored in the inductor is in the form of kinetic energy of electrons. 
It is not in the form of potential energy. 
This can be proven by complex maths that kinetic energy of electrons moving with drift velocity 
in a solenoid is exactly equal to the magnetic energy.\rq\rq\ 
 \cite{stackexchange}

\newpage
\section*{Post Scriptum}

After having derived the formula for the
self-inductance of an ideal solenoid,
and after a brief moment of euphoria,
some doubts have crept into my mind.
How could such an important result not be part of the standard literature on the subject?
Could it be that I have missed some relevant bibliographic references? 

I realized that ChatGPT offers a very efficient way of indirectly examining all the published information
on a given topic,
so I asked the question: \lq\lq What is the relationship between the magnetic energy stored in a coil and the 
kinetic energy of the electrons?\rq\rq\ 
Among other things, the answer (from November 2024) said:
\lq\lq Electrons moving through the coil have a small kinetic energy due to their motion. 
However, in typical electrical circuits, this kinetic energy is negligible compared to 
the total energy stored in the magnetic field. [...]
In conductors, drift velocity $v_d$ is quite small, so this kinetic energy is minimal. [...]
The magnetic energy stored in the coil comes from the collective effect of the movement of many electrons 
rather than their individual kinetic energies.\rq\rq\ 

Soon after reading this unforgiving ChatGPT answer 
I found the same conclusion in the comprehensive analysis of Kirk~T.~McDonald \cite{kirkmcdonald},
published on his website among other valuable physics essays,
where he explains why the hypothesis put forward by E.~G.~Cullwick is \lq\lq unsupportable\rq\rq.

The next thing I discovered was the mistake in my own derivation of formula (\ref{eq34}).
It turns out that the magnetic field created by the moving conduction electrons points into the opposite direction.
In other words, the magnetic force acting on the moving electrons points outward, 
away from the centre of the circular loop of wire.
This magnetic force alone cannot be responsible for the helicoidal motion of the electrons.

The logical conclusion was that there must be a radial electric field inside the wire, responsible for the centripetal force.
This is in stark contrast with the previously quoted statement 
from the 15-th edition of {\it Sears and Zemansky} \cite{zemansky}.
I took their words for granted, without proper justification, and this has sent me on the wrong track.
The total electric field $\vec{E}_c + \vec{E}_n$ within the coils is not zero, 
but must consist of a radial electric field $\vec{E}_r$ that is perpendicular to the tangent to the wire.
In this way, when $\vec{E}_c + \vec{E}_n = \vec{E}_r$ and $\vec{E}_r \cdot \vec{dl} = 0$, 
the calculation from Eq. (\ref{eq4}) still gives the same result.

In fact, with or even without a magnetic field, such an internal electric field perpendicular to the drift velocity 
must be present whenever the wire bends, in order to guide the motion of the conduction electrons.
There must be some accumulations of electric charge on the surface of a curved wire,
producing a transverse electric field and a transverse voltage in the plane of the cross-section of the wire,
that are in complete analogy with those seen in the Hall effect. 
This non-magnetic effect has been investigated theoretically and experimentally by
Nicholas B. Schadea, David I. Schuster, and Sidney R. Nagel,
who have observed transverse voltages as large as millivolts
in curved graphene wires. 
\lq\lq In the classical Hall effect,
a magnetic field curves the paths of charge carriers inside a
straight wire so that charges accumulate on the wire edges
transverse to the current. In the geometric analog, we do not
bend the paths of the carriers but instead bend the conductor
itself to create a purely geometric effect.\rq\rq\ \cite{pnas}

We can get closer to the correct description of the inertial effect 
experienced by an electric current flowing in a solenoid 
with the help of the Lagrangian formalism. 
The Lagrangian of an electrically charged particle in a stationary
electromagnetic field is
\begin{equation}
L = \frac{1}{2} m \, \vec{v}^{\, 2} - q \, \phi + q \, \vec{v} \cdot \vec{A},
\end{equation}
where $\phi(x, y, z)$ is the electric (scalar) potential 
and $\vec{A}(x, y, z)$ is the magnetic (vector) potential
at the position of the particle. 
In the absence of an electromagnetic field, $\phi$ and $\vec{A}$ have null values everywhere.
The $x$, $y$, and $z$ position coordinates are cyclic coordinates,
and the canonical momentum, which in this case is equal to 
the mechanical momentum $\vec{p} = m \, \vec{v}$,
is constant. This is Newton's first law of motion.
However, with a magnetic field present, the canonical momentum 
becomes 
\begin{equation}
\vec{P} = m \, \vec{v} + q \, \vec{A},
\end{equation} 
where the last term is the electromagnetic \lq\lq potential\rq\rq\ momentum \cite{GriffithsAJP2012}.
Due to the symmetry of the problem, we switch to cylindrical coordinates $r$, $\theta$, $z$.
For a stationary electric field in the radial direction, of constant magnitude for any $\theta$ or $z$, the scalar potential is $\phi(r)$.
For an infinitely long ideal solenoid of radius $R$ that has 
a stationary magnetic field $B_z = B$ inside and $B_z = 0$ outside, 
the vector potential is $A_\theta(r) = B \, r / 2$ inside and $A_\theta(r) = B \, R^2 / (2 \, r)$ outside.
The Lagrangian in cylindrical coordinates becomes
\begin{equation}
L = \frac{1}{2} m \left(\dot{r}^2 + r^2 \, \dot{\theta}^2 + \dot{z}^2\right) - q \, \phi(r) + q \, r \, \dot{\theta} \, A_\theta(r).
\end{equation}

The radial electric field compels the conduction electrons to follow the turns of the wire, 
thus $r = R$ and $\dot{r} = 0$.
Since $A_\theta(r)$ is continuous but not differentiable when $r = R$,
the value of $\partial{\phi}/\partial{r}$ when $r = R$ cannot be extracted from 
the corresponding Euler-Lagrange equation, and this is a deficiency of the model. 
The $z$ and $\theta$ generalized position coordinates are cyclic coordinates, and the canonical momenta 
$m \, \dot{z}$ and $m \, r^2 \ \dot{\theta} - q \, r \, A_\theta(r)$ are conserved.

Due to the fact that the electrons are forced to move along the filamentary wire,
which allows for only one degree of freedom,
if one component of their velocity is determined, 
{\it e.g.} $v_\theta = R \, \dot{\theta}$,
then the other components are also determined, 
{\it e.g.} $\dot{r} = 0$ and $\dot{z} = \ell / (N \, T) = \ell \, \dot{\theta} / (2 \pi \, N)$.
And due to the assumed translational and rotational symmetry 
along the $z$ axis of the infinitely long ideal solenoid, 
if the speed of one conduction electron is determined,
then this must also be the speed of any other conduction electron.

Now we begin to understand \lq\lq the collective effect of the movement of many electrons\rq\rq. 
At the position of each conduction electron, 
the angular component $A_\theta(R)$ of the vector potential,
the only non-zero component, is due to the 
retarded position, the retarded velocity, and the electric charge of all the other conduction electrons.
The superposition of all these Li\'{e}nard-Wiechert magnetic potentials 
produces a relationship between $A_\theta(R)$
and the angular velocity $\dot{\theta}$,
which is the same for all the conduction electrons. 
Since the conservation
of the canonical momentum associated with the angular variable $\theta$ provides a second equation, 
we now have a system of two equations with two unknowns. 
This fixes the values for both $A_\theta(R)$ and $\dot{\theta}$.
In the absence of an external voltage source or sink (electrical resistance), 
the electric current in the infinitely long ideal solenoid will flow forever with constant intensity.

For an induced voltage (e.m.f.) to appear, the translational symmetry of the system must be broken,
and in the most general case the electric potential $\phi(r, z, t)$ will also change with the time.
In the non-stationary case, for the calculation of the electric and magnetic fields,
the Coulomb and the Biot-Savart laws must be replaced with 
Jefimenko's equations \cite{GriffithsHealdAJP1991}.
The magnetic field will no longer be parallel to the solenoid axis, 
or permanently restricted to the spatial region inside the windings,
and this will allow the Poynting vector to bring energy into the solenoid \cite{KirkTMcDonaldAJP1997}.
It is beyond the scope of this study to pursue such calculations.

Even though my naive derivation of the formula for the self-inductance of an ideal solenoid 
proved to be incorrect, I still believe that this derivation has great intrinsic educational value.
It teaches the lesson that one can derive a correct physics formula based on the wrong assumptions.
It teaches the lesson that one has to clearly and carefully 
identify all the hypotheses and assumptions involved,
and has to constantly reassess their validity, always and without exceptions.
On this occasion it came to light that some of the information presented in two 
textbooks \cite{cullwick, zemansky},
one written by a Professor of Electrical Engineering, also a Fellow of the Royal Society of Edinburgh, 
and the other one among the most widely used introductory physics textbooks in the world, was wrong.
Not only this, but some of the information presented in an article \cite{heald} 
published in the {\it American Journal of Physics} was also wrong.

It is important to learn physics, but it is also important to learn about how we learn physics, 
and in particular about how we deal with longstanding misconceptions and misrepresentations, 
untenable hypotheses, and failures of our intuition. 
The incorrect derivation presented here can provide a moment of individual introspection and enlightenment
that can lead to improved critical thinking and true physics mastery. 
I am convinced that all those who have unwittingly fallen into the trap this time,
believing even for a brief moment that the mechanical inertia of the conduction electrons 
is responsible for the electrical inertia (the self-inductance) of a solenoid,
will have a good laugh, will forgive me for this mischievous arXiv contribution, 
and will be more aware of the \lq\lq dangers\rq\rq\ lurking in physics next time.

\end{document}